\documentclass[11pt]{article}
\usepackage[utf8]{inputenc}
\usepackage[margin=1in]{geometry}
\usepackage{amsmath}
\usepackage{amsfonts}
\usepackage{amssymb}
\usepackage{url}
\usepackage{hyperref}
\usepackage{graphicx}
\usepackage{xcolor}
\usepackage[most]{tcolorbox}
\usepackage{fancyvrb}
\usepackage{setspace}
\usepackage{titlesec}

\titleformat{\section}
{\normalfont\Large\bfseries}{\thesection}{1em}{}
\titlespacing*{\section}{0pt}{18pt plus 4pt minus 2pt}{12pt plus 2pt minus 2pt}

\titleformat{\subsection}
{\normalfont\large\bfseries}{\thesubsection}{1em}{}
\titlespacing*{\subsection}{0pt}{14pt plus 3pt minus 2pt}{8pt plus 2pt minus 2pt}

\titleformat{\subsubsection}
{\normalfont\normalsize\bfseries}{\thesubsubsection}{1em}{}
\titlespacing*{\subsubsection}{0pt}{12pt plus 2pt minus 2pt}{6pt plus 2pt minus 2pt}

\newtcolorbox{protocolbox}{
  colback=gray!5,
  colframe=gray!30,
  boxrule=0.8pt,
  arc=4pt,
  left=12pt,
  right=12pt,
  top=12pt,
  bottom=12pt,
  before skip=12pt,
  after skip=12pt
}

\DefineVerbatimEnvironment{protocolcode}{Verbatim}{
  fontsize=\footnotesize,
  fontfamily=tt,
  formatcom=\color{black},
  baselinestretch=1.2,
  commandchars=\\\{\},
  samepage=true
}

\fvset{listparameters=\setlength{\topsep}{6pt}\setlength{\partopsep}{0pt}}

\title{The Stated Protocol: A Decentralized Framework for Digital Diplomacy}

\author{
Christopher J. P. Rieckmann\\
\texttt{cjp.rieckmann@gmail.com}\\
\texttt{globalcoordination.org}
}

\date{\today}

\begin{document}

\maketitle

\begin{abstract}
\setstretch{1.2}
International coordination faces significant friction due to reliance on
periodic summits, bilateral consultations, and fragmented communication channels
that impede rapid collective responses to emerging global challenges while
limiting transparency to constituents.

We present the Stated Protocol, a decentralized framework that enables
organizations to coordinate through standardized text statements published on
their website domains. While applicable to all organizations, this work focuses
primarily on the application in international relations, where the protocol
enables rapid consensus discovery and collective decision-making without relying
on centralized social media platforms.

We explore specific applications: (1) faster treaty negotiation through
incremental micro-agreements that can be signed digitally within hours rather
than months, (2) continuous and transparent operation of international
institutions through asynchronous decision-making, (3) coordinated signaling
from local governments to national authorities through simultaneous statement
publication, and (4) coalition formation among non-governmental organizations
through transparent position aggregation.
\end{abstract}

\vspace{12pt}

\section{Introduction}
Many preventable global crises - from ongoing armed conflicts to climate change
and nuclear proliferation - remain unresolved despite widespread recognition of
their urgency. These failures may stem from insufficient political will, or from
ineffective coordination mechanisms preventing latent political will from being
effectively expressed and aggregated into collective action.

Current international coordination often relies on fragmented, slow-moving
processes that create systematic barriers to rapid responses. Ineffective
coordination makes international law enforcement unreliable, which in turn
reduces political will by states to engage in binding agreements - if the terms
may not be enforced on their counterparts, creating a self-reinforcing cycle of
international diplomatic ineffectiveness.

Coordination friction also creates asymmetric disadvantages for smaller nations,
who face disproportionate risks when negotiating with larger powers in isolation
- potentially further decreasing their political will to engage with global
challenges. Without mechanisms for rapid coalition formation, smaller states
cannot effectively aggregate their collective influence, leaving them vulnerable
to bilateral pressure and reducing their ability to shape global outcomes.

Effective coordination frameworks need to enable fast and low-cost discovery of
shared positions and fast and low-cost creation of collective agreements.
Current diplomatic methods often rely on outdated traditions and appear to
impede rapid collective responses particularly to complex issues. This work
focuses on coordination scenarios where transparency is permissible,
complementing existing confidential diplomatic channels.

\subsection{Challenges in Current Diplomacy Practices}
International communication appears to rely on fragmented approaches: numerous
bilateral conversations, scattered press releases from foreign ministries,
in-person voting by delegates at periodic international meetings, and ad-hoc
communication through various social media platforms.

\begin{figure}[h]
\centering
\begin{tcolorbox}[colback=white, colframe=gray!50, boxrule=1pt, arc=3pt,
left=3pt, right=3pt, top=3pt, bottom=3pt, width=0.7\textwidth, drop shadow
southeast]
\includegraphics[width=\linewidth]{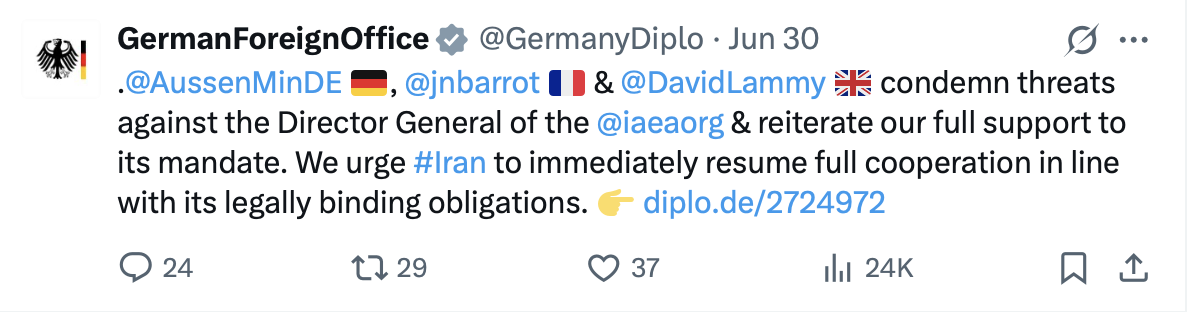}
\end{tcolorbox}
\caption{German Foreign Ministry X/Twitter post demonstrating current coordination challenges in international diplomacy}
\label{fig:xpost}
\end{figure}
A recent example illustrates coordination limitations: On June 30, 2025, the
German Foreign Ministry posted on X (formerly Twitter) condemning threats
against the IAEA Director General, tagging French and British foreign ministers
and linking to a German press release (see Figure~\ref{fig:xpost}). While this
post suggests agreement, it is not immediately obvious whether France and
Britain issued identical statements via official channels, what the exact
content alignment is across the three nations, or what positions the remaining
177 IAEA member states hold on this issue. Within x.com, other nations do not
have a suitable option to add their position to the post after it has been
published.

\begin{figure}[h]
\centering
\begin{tcolorbox}[colback=white, colframe=gray!50, boxrule=1pt, arc=3pt,
left=3pt, right=3pt, top=3pt, bottom=3pt, width=0.7\textwidth, drop shadow
southeast]
\includegraphics[width=\linewidth]{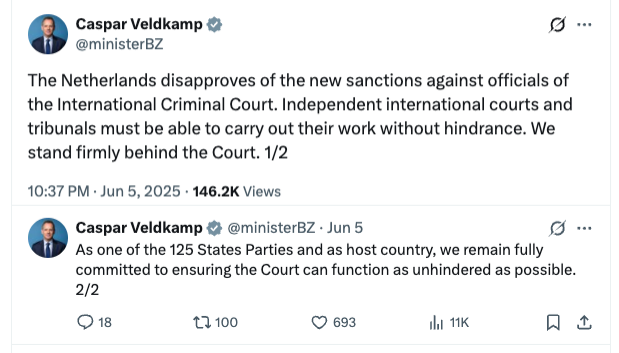}
\end{tcolorbox}
\caption{Dutch Foreign Ministry X/Twitter post on ICC sanctions, demonstrating similar coordination opacity in international responses}
\label{fig:xpost_icc}
\end{figure}
Another example from June 5, 2025, shows similar limitations: The Dutch Foreign
Ministry posted on X expressing disapproval of new US sanctions against
International Criminal Court employees and their immediate family members
\cite{whitehouse2025icc, state2025icc}, stating their firm support for the Court
(see Figure~\ref{fig:xpost_icc}). However, this post lacks the coordination
mechanisms essential for actionable collective response - clear deadlines,
enforcement measures, and pathways for other nations to formally align their
positions. Apparently, no widely used framework exists for the other 124 ICC
states to systematically coordinate a rapid response and establish appropriate
collective enforcement mechanisms. Without standardized coordination protocols,
such diplomatic expressions remain isolated statements rather than catalysts for
effective international action.

\subsubsection{Scalability Limitations}
Some of these traditional coordination modes exhibit fundamental scalability
constraints. Bilateral coordination scales quadratically - for 100 nations,
there are 4,950 bilateral communication channels, rendering comprehensive
consultation impractical at this scale. In-person meetings, typically granting
each member speaking time, become prohibitively time-intensive as membership
expands. Delegate-based coordination requires frequent consultation delays with
home governments due to limited negotiation power and information, while the
delegate interaction may be inaccessible to the public. Fragmented press
releases across different platforms and formats prevent automated position
aggregation and real-time consensus discovery, forcing diplomats to manually
track and interpret scattered communications across heterogeneous information
sources.

\subsubsection{Hierarchical Coordination}
Static group formation among countries with correlated interests, such as the
BRICS and G7, allows countries to align among their members while avoiding
scalability limitations. This allows them to negotiate as a bloc in global
forums and UN voting. However, on a global level, coordination may be worsened
by group alignment delays and polarization.

\subsubsection{Enforcement Challenges}
International rule enforcement compounds these coordination challenges. Absent
supranational police enforcement, compliance depends commonly on collective
economic sanctions that require broad participation to be effective. The
existence of even a single sanction evasion hub country can severely limit the
effectiveness of such sanctions.

\subsubsection{Platform Dependency Vulnerabilities}
Beyond coordination inefficiencies, reliance on centralized social media
platforms introduces systemic vulnerabilities to government communications. In
2009, Facebook removed Macedonia's Ministry of the Interior's profile,
demonstrating how centralized platforms can unilaterally disrupt official
government communications \cite{stojanovski2009}. Another notable example
occurred in 2021 when Twitter verified a fake account impersonating Norway's
Minister of Finance, despite the minister never having a Twitter account,
demonstrating the vulnerability of opaque centralized verification systems even
for high-profile government officials \cite{roth2021}.

\section{The Stated Protocol}

\subsection{Overview}
The Stated Protocol reduces coordination friction among organizations through a
decentralized framework that leverages existing web infrastructure.
Organizations publish standardized text statements at
\texttt{/.well-known/\allowbreak statements.txt} on their domains, enabling
automated position aggregation, rapid coalition formation, and digital treaty
signing without relying on centralized platforms.

\subsection{Statement Structure}
Each statement follows a standardized format with required and optional fields:

\begin{protocolbox}
\begin{protocolcode}
Publishing domain: <domain>
Author: <organization>
Authorized signing representative: <representative> (optional)
Time: <UTC timestamp>
Tags: <comma-separated tags> (optional)
Superseded statement: <base64-encoded-SHA256-hash> (optional)
Format version: 4
Statement content: <content or typed statement>
\end{protocolcode}
\end{protocolbox}

\textbf{Design Principles:} The format deliberately avoids complex syntax,
special characters, or technical jargon that might create barriers to adoption
or comprehension. Field names like \texttt{Publishing domain} and
\texttt{Author} are immediately understandable, while the content uses natural
language constructs such as \texttt{Type: Poll} and \texttt{Option 1: Yes}
rather than encoded identifiers. Reduced degrees of freedom, such as the fixed
field sequence, ensure reproducible statement content hashes which facilitate
position aggregation via hashes.

\textbf{Technical Requirements:} All statement files must use UTF-8 encoding
without BOM (Byte Order Mark). Servers must serve statement files with the HTTP
Content-Type header \texttt{text/plain; charset=utf-8} to ensure proper
interpretation by clients. This ensures consistent interpretation across
different systems and platforms while supporting international characters in
organization names, content, and metadata.

\textbf{File Organization:} Statements are separated by double newlines and
served as \texttt{text/plain; charset=utf-8}. Individual statements can be
accessed via \texttt{/.well-known/\allowbreak statements/\allowbreak <hash>.txt}
where the hash is the URL-safe base64 encoding of the SHA-256 hash of the
statement.

\subsection{Statement Types}
The protocol supports multiple statement types organized into three functional
categories. All typed statements are nested inside the \texttt{Statement
content:} field with tab indentation, with each field-content pair separated by
a newline.

\textbf{Example Statement Structure:}
\begin{protocolbox}
\begin{protocolcode}
Publishing domain: example.gov
Author: Ministry of Foreign Affairs
Time: 2027-01-01T10:30:00Z
Format version: 4
Statement content:
	Type: Sign PDF
	Description: We hereby digitally sign the referenced PDF file.
	PDF file hash: qg51IiW3RKIXSxiaF_hVQdZdtHzKsU4YePxFuZ2YVtQ
\end{protocolcode}
\end{protocolbox}
For detailed technical specifications, see \url{https://stated.network/docs/stated_well_known.html}.

\subsubsection{Identity Establishment}
These statement types enable organizations to verify identities and build trust
networks:

\textbf{Organisation Verification:} Enables organizations to verify other
organizations' online identities, creating a web of trust for institutional
identity. Organizations provide confidence ratings (0.0-1.0) representing their
certainty in the verification:

\begin{protocolbox}
\begin{protocolcode}
	Type: Organisation verification
	Description: We verified the following information about an organisation.
	Name: <organization name>
	Country: <country>
	Legal form: <legal form>
	Owner of the domain: <domain>
	Confidence: <0.0-1.0>
\end{protocolcode}
\end{protocolbox}

\textbf{Person Verification:} Similar verification capabilities for individual
identity attestation, requiring name, birth details, and domain ownership
verification with confidence ratings.

\subsubsection{Coordination and Decision-Making}
These statement types facilitate collective actions, from simple communication
to complex treaty negotiations:

\textbf{Plain Content:} Unstructured text statements for general organizational
communication, comparable to institutional press releases or social media posts.
This provides the foundation for transparent organizational communication.

\textbf{Sign PDF:} Enables collective document endorsement through cryptographic
hash references, allowing multiple organizations to digitally sign shared
agreements. The referenced PDF documents should be published under
\texttt{/files/\allowbreak <hash>.pdf}, creating verifiable treaty signatures
and policy endorsements.

\textbf{Poll:} Creates structured voting opportunities with defined options,
deadlines, and eligibility criteria for collective decision-making:
\begin{protocolbox}
\begin{protocolcode}
	Type: Poll
	Voting deadline: <UTC timestamp>
	Poll: <question>
	Option 1: <choice>
	Option 2: <choice>
	Who can vote:
		Description: <eligibility criteria>
\end{protocolcode}
\end{protocolbox}

\textbf{Vote:} Enables participation in existing polls by referencing the poll
hash and selected option.

\textbf{Response:} Facilitates threaded conversations by referencing existing
statements via their hash, enabling structured organizational dialogue and
building upon previous communications.

\textbf{Bounty:} Allows organizations to offer rewards for specific actions or
achievements, facilitating incentive-based coordination.

\textbf{Boycott:} Facilitates coordinated economic pressure campaigns against
specific subjects, providing a mechanism for collective action and enforcement.
For the specification of complex terms, signing a PDF statement may be
preferred.

\subsubsection{Content Moderation and Quality Control}
These statement types enable crowdsourced content verification and decentralized
accountability frameworks:

\textbf{Dispute Authenticity:} Challenges the authenticity of statements,
enabling network participants to flag potential impersonation attempts and
maintain network integrity.

\textbf{Dispute Content:} Challenges the factual accuracy of statement content
by referencing the statement's hash, creating a mechanism for fact-checking and
content verification.

\textbf{Rating:} Rates an organization on defined qualities such as
trustworthiness using a 1-5 star scale, building reputation systems that inform
network participants about organizational credibility.

\section{Applications}
The Stated Protocol enables a number of capabilities for inter-organizational
coordination across diverse contexts:

\textbf{Platform-Independence:} Organizations maintain full control over
publishing statements without relying on centralized platforms or single
jurisdictions.

\textbf{Rapid Coalition Formation:} Automated aggregation of identical
statements enables efficient consensus discovery.

\textbf{Collective Document Signing:} The generic document signing solution
enables a range of actions, such as the creation of joint statements, collective
contracts, international treaties or narrowly scoped micro-agreements.

\textbf{Strong Identities:} Transparent and numerous independent identity
verifications among participating organizations can constitute a high level of
aggregated confidence in the online identities. Strong identities are necessary
for quickly establishing consequential collective actions as widely regarded
truths.

\subsection{International Diplomacy}

\subsubsection{Efficiency in Position Discovery}
Traditional diplomacy requires discovering each country's position through
separate press releases or bilateral consultation, with time investment scaling
linearly with participating countries ($n_{countries} \times t_{per\_country}$).
The protocol enables automatic aggregation of published positions, where
discovery time depends only on the number of unique negotiating positions
($n_{unique\_positions} \times t_{per\_position}$) - typically far fewer than
participating countries.

\subsubsection{Strategic Advantages for Smaller Nations}
The protocol provides particular benefits for smaller nations by reducing
coordination friction in coalition building. In traditional diplomacy, smaller
states face asymmetric power dynamics where larger powers can exploit high
coordination costs. The Stated Protocol enables rapid consensus signaling among
collections of smaller countries, potentially creating more visible collective
positions that increase their aggregate negotiating power.

Multiagent coordination simulations suggest that communication friction delays
coordination onset and may prevent coordination entirely
\cite{herreramedina2023}. By reducing coordination friction, the protocol may
contribute to more balanced negotiation dynamics, enabling smaller states to
signal collective positions more rapidly than traditional methods allow.

\subsubsection{Transparency and Accountability}
Public, independently verifiable statements facilitate constituent scrutiny and
promote positions defendable to global citizens rather than special interests,
countering targeted manipulation, over-reliance on personal connections and
misrepresentation.

\subsubsection{Document Signing and Treaties}
The EU's Lisbon Treaty exemplifies some problems with traditional approaches:
large texts edited by multiple parties in an opaque process led to the inclusion
of provisions that none of the member states actually intended
\cite{adlernissen2019}.
The Stated framework facilitates transparent treaty development through public,
threaded discussions wherein nations can engage in structured discourse
referencing specific treaty draft clauses. Through the Response statement type,
participating states can systematically address particular provisions, propose
amendments, and build consensus around treaty language before formal adoption.
This mechanism enables iterative refinement of treaty texts through documented
exchanges that reference specific sections. Traditional treaty creation bound to
infrequent in-person meetings incentivizes consolidating disparate issues into
complex documents, diluting focus and accountability. The protocol's reduced
friction enables the creation of smaller, more focused, and more frequent
treaties.

The ease of creating agreements asynchronously may enable new diplomatic
approaches, such as iterative, incremental position-taking where nations engage
in multiple successive micro-commitments contingent on coalition support.

\subsection{Local-to-National Government Signaling}
The protocol enables local governments to signal consensus on underappreciated
problems to national authorities through coordinated statement publication. When
municipal and regional authorities collectively publish statements on specific
issues, they create visible, quantifiable evidence of widespread local concern
that may otherwise be overlooked by national decision-makers.

\subsubsection{Bypassing Hierarchical Communication}
This signaling mechanism proves particularly valuable in multi-layered
representative systems, such as European institutions where multiple layers of
representatives can create distance between citizen concerns and policy
decisions. Local governments can bypass traditional hierarchical communication
channels and lobbying groups by directly publishing coordinated statements that
demonstrate clear consensus on issues requiring national or supranational
attention.

\subsubsection{Applications Across Political Systems}
In democratic systems, the protocol strengthens representative democracy by
creating more responsive feedback loops between different levels of government,
ensuring that local-level consensus reaches decision-makers who might otherwise
remain unaware of grassroots priorities.

In authoritarian regimes, where power centers are often insulated from
grassroots concerns, rapidly coordinated signaling from local governments may
provide the amplification needed to reach decision-makers. The protocol's
transparency and verifiability help document widespread local consensus,
potentially creating accountability pressure through documented, attributable
statements from legitimate governmental entities.

\subsection{Digital International Institutions}
The Stated Protocol offers building blocks for new digital-first international
institution structures with improved coordination mechanisms that could serve as
digital alternatives or complements to organizations like the United Nations.

\subsubsection{Continuous Operation and Rapid Response}
Digital institutions built on the Stated Protocol could operate continuously
rather than through periodic summits or in-person representatives, could respond
rapidly to emerging crises, and could maintain transparent, easily accessible
records of member positions and commitments.

\subsubsection{Enhanced Rule Enforcement}
Enabling nations to negotiate and sign documents with minimal effort could
improve international rule enforcement by enabling rapid creation of detailed
punishment agreements in response to even small transgressions. These agreements
could outline precise schedules of ratcheting sanctions, reversal conditions,
and transparent escalation procedures, creating more effective deterrence
mechanisms.

\subsection{Coalition Formation Among Non-Governmental Organizations}

\subsubsection{Discovering and Demonstrating Alignment}
The protocol's reduced coordination friction enables organizations to discover
and demonstrate stance alignment on nuanced topics that might otherwise remain
uncommunicated or fragmented across individual posts or press releases. This
capability proves particularly valuable for organizations seeking to coordinate
responses to complex policy issues that affect multiple sectors.

\subsubsection{Countering Special Interest Influence}
The protocol empowers organizations to assume coordinated counter-positions
against special interest representation. Collective voice amplification can
counterbalance well-funded advocacy efforts that seek disproportionate benefits
for small groups at the expense of broader societal interests.

\subsubsection{Low Barriers to Coalition Participation}
The protocol's low participation barriers enable organizations that were not
specifically formed to represent certain interests to spontaneously join
transnational coalitions when issues align with their broader missions.
Universities, professional associations, environmental groups, and other
organizations can easily participate in coordinated responses without requiring
dedicated advocacy or lobbying setups, expanding potential coalition size beyond
traditional opposition groups.

\section{Security Analysis}
The Stated Protocol's security architecture relies on the internet's established
DNS and TLS infrastructure, creating a robust foundation for organizational
communication. Statement authenticity is verified through domain ownership,
requiring attackers to compromise DNS records, TLS certificates, or the
organization's website infrastructure to forge statements. The reference
implementation also supports publication of statement hashes as DNS TXT entries
to mitigate website compromise attacks. The decentralized nature reduces single
points of failure compared to centralized platforms while maintaining verifiable
attribution.

This domain-based trust model builds on established principles from federated
identity management systems, where domains serve as trust anchors for
organizational identity \cite{chadwick2009}. The approach leverages existing
security investments while adding protocol-specific verification mechanisms.

\subsection{Web of Trust}
Trust and acceptance are critical variables in diplomatic technology adoption
\cite{almuftah2018}. Foreign ministries and other organizations possess
well-known domains with existing reputational stakes, creating natural trust
anchors for the verification network.

\subsubsection{Confidence-Based Verification System}
The verification system operates at the organizational level with quantitative
confidence ratings (0.0-1.0), unlike PGP's binary trust decisions
\cite{ulrich2011}. The confidence rating represents a guaranteed probability of
correctness, with verifiers committing to pay bounties if confidence levels are
statistically disproven, creating economic incentives for accurate assessments
along with well-calibrated confidence levels.

For independent verifications with confidences $c_1, c_2, \ldots, c_n$, the
combined confidence equals: $$C_{aggregate} = 1 - \prod_{i=1}^{n}(1 - c_i)$$

For example, three independent verifications at 0.8 confidence yield:
$$C_{aggregate} = 1 - (1-0.8)^3 = 0.992$$

This example demonstrates how distributed verification can achieve
high-confidence identity attestation comparable to centralized certificate
authorities, provided that the judgments are based on independent data and
independent verification processes.

\subsection{Threat Model}
The protocol faces four primary threat categories, each with specific mitigation
strategies:

\subsubsection{Domain Compromise}
Attackers gaining control of organizational domains can publish false
statements. Current mitigation relies on \texttt{Dispute statement authenticity}
and \texttt{Dispute statement content} statements that enable organizational
verification networks to flag suspicious behavior patterns. Additionally,
legitimate organizations can use the \texttt{superseded statement} mechanism to
revoke compromised statements by publishing new statements that reference and
invalidate fraudulent ones.

Future enhancements could include a separate Public Key Infrastructure layer
with hardware-controlled signing keys (e.g., foreign minister-controlled
hardware) that cryptographically sign official statements, providing additional
authentication independent of DNS/TLS infrastructure.

\subsubsection{Impersonation}
Malicious actors may register domains that resemble legitimate organizations and
attempt impersonation. The web of trust system mitigates this risk through
organizational verification networks that can identify and flag suspicious
impersonation attempts, leveraging the collective intelligence of verified
organizations.

\subsubsection{Coordination Attacks}
For the non-diplomatic use cases, malicious actors may try to create large
numbers of fake organization identities to fabricate support for certain
positions, potentially creating verification networks among these fake
organizations. This may be mitigated by analyzing verification network topology
for suspicious clustering patterns, and cross-referencing with established
databases and pre-existing SSL organization validation certificates.

\subsubsection{Censorship}
Nations may attempt to censor certain organizations by blocking IP addresses or
DNS queries. To successfully censor collective actions, a large fraction of the
participating organizations would have to be blocked simultaneously. The
protocol's decentralized architecture also enables mitigation via propagation
and aggregation of statements across a P2P network, as it is done in the
referenced implementation, described in the following section.

\section{Implementation}
A complete reference implementation of the Stated Protocol has been developed in
TypeScript, demonstrating the protocol's practical feasibility. The
implementation is available at:

\url{https://github.com/c-riq/stated}
The system consists of three main components that work together to provide a
complete coordination platform:

\subsection{Statement Generation UI}
A web interface that allows organizations to create and publish statements
across all statement types including polls, votes, document signatures, and
organizational verifications.

\subsection{P2P Network Nodes}
Distributed Node.js PostgreSQL nodes that broadcast, validate, and aggregate
statements across the network. The nodes use a pull gossip protocol for
synchronization, assigning incrementing statement ID integers to track
synchronization state. Nodes periodically pull updates from a random subset of
peer nodes and deduplicate statements using SHA-256 hashes to ensure eventual
consistency without requiring consensus mechanisms. To mitigate spam and ensure
network quality, a reputation score is maintained for each node.

\subsection{Aggregation Display UI}
A social-media-like interface that shows aggregated results, allowing users to
browse statements, view consensus positions, track voting results, and inspect
verification relationships. The system integrates with SSL OV (Organization
Validation) certificates to bootstrap organizational online identities,
leveraging existing Certificate Authority infrastructure while enabling the
transition to an independent peer-to-peer verification network.

\section{Adoption Considerations}
Al-Muftah et al. identify resistance to change as a critical barrier in digital
diplomacy adoption \cite{almuftah2018}. The protocol addresses adoption barriers
through minimal technical requirements: organizations need only publish a plain
text file at \texttt{/.well-known/\allowbreak statements.txt} on their existing
websites - no specialized software installation, private key generation, or
platform registration required. This low barrier to entry facilitates adoption
across diverse organizational types and technical capabilities.

\section{Limitations and Scope}

\subsection{Personal Use Constraints}
While the protocol supports both organizational and individual participation,
direct individual use faces two fundamental barriers that limit its
applicability for personal coordination: (1) Most individuals do not own website
domains required for statement publication, creating a significant barrier to
direct participation in the protocol's coordination mechanisms. (2) Individual
participation requires identity verification to prevent duplicate participation,
which necessitates revealing sufficient data points to uniquely identify a
person. This requirement conflicts with privacy norms. However, individual
participation becomes feasible when organizations act as intermediaries. Local
governments, employers, universities, or other institutions could serve as
verification authorities and statement publishers, effectively addressing both
barriers and enabling broader participation beyond organizational use.

\subsection{Confidential Diplomacy}
The protocol is explicitly designed for transparent coordination and does not
address legitimate confidentiality requirements in diplomatic practice.
Sensitive negotiations involving security intelligence or trade secrets that
require confidential communication channels fall outside the protocol's intended
scope. The framework complements rather than replaces existing confidential
diplomatic mechanisms, focusing specifically on coordination scenarios where
transparency enhances rather than impedes effective collective action.

\subsection{Scalability Considerations}
For international coordination among approximately 200 nations publishing
statements on their websites with fewer than 1000 independent aggregator nodes
polling these sites, scalability concerns are minimal given current internet
infrastructure capabilities.

However, large-scale adoption among numerous local governments, NGOs, and
companies would benefit from specialized P2P network nodes that focus on
specific organizational types. This specialization would prevent redundant
polling of all publishing websites by all aggregators, reducing overall network
load and improving system efficiency.

\section{Related Work}
The protocol builds on established web standards including RFC 5785 (well-known
URIs), ISO 8601 (timestamp format), and existing DNS/TLS infrastructure. While
individual components exist in isolation, no existing system combines
decentralized domain-based publishing, human-readable protocol close to natural
language, and a confidence-based web of trust verification to enhance
organizational collective action capabilities.

It should be noted that there is no consensus formation required among the
stated network nodes. Therefore, blockchain mechanisms are not involved in the
realization.

\subsection{Comparison with ActivityPub}
ActivityPub \cite{activitypub2018} and the Stated Protocol share a key
commonality as generic frameworks for decentralized interactions. ActivityPub
provides a generic social networking protocol supporting various activity types
(Create Note, Like, Follow ...), with Mastodon being one of the major
decentralized social networks implementing this protocol
\cite{mastodon2024}. The Stated Protocol provides a generic institutional
coordination framework supporting different types, which are in part overlapping
(Organisation Verification, Sign PDF ...).

In addition to the different activity types resulting from differing target
groups and use cases, they diverge fundamentally in two key areas:

\textbf{Data Format:} ActivityPub uses JSON-LD with ActivityStreams vocabulary -
a machine-readable format with complex structure and extensive metadata. The
Stated Protocol uses intuitive English language field descriptors separated from
field values by a colon, deliberately avoiding technical jargon and obscure
special characters to remain close to natural language. This enables managers
and government officials who may not be familiar with the detailed specification
to independently verify the publication of a statement. In addition,
ActivityPub, extending on JSON, does not rely on key-value pair order. The
Stated Protocol, however, enforces a field order and thereby ensures
reproducible statement content hashes - resulting from identical character
sequences - which is essential for hash based aggregation.

\textbf{Communication Architecture:} ActivityPub implements real-time push-based
federation where servers POST activities to each other's inboxes for immediate
delivery and live social interaction. The Stated Protocol supports asynchronous
pull-based communication through domain-based publishing at
\texttt{/.well-known/\allowbreak statements.txt} files with static file serving.

\section{Future Work}
Future research could evaluate framework adoption impact through pilot
deployments with specific government ministries and measure coordination time
improvements compared to alternative diplomatic channels.

Key areas for development include:

\setlength{\itemsep}{8pt}

\textbf{Tooling Development:} Organizations may benefit from simplified
publishing tools and website management software integrations to ease
decentralized statement publication.

\textbf{Protocol Extensions:} Real-world usage may identify missing statement
types and necessary protocol modifications to better serve coordination needs.

\section{Conclusion}
The Stated Protocol addresses fundamental coordination challenges through a
decentralized framework that leverages existing web infrastructure. The
framework is immediately deployable - requiring only plain text file publication
on existing domains - which minimizes technical barriers to adoption while
maintaining human readability and automated processing capabilities.

Applications spanning international diplomacy, local-to-national government
signaling, digital international institutions, and NGO coordination demonstrate
the protocol's versatility across diverse organizational contexts.

The protocol's foundation in established web standards and demonstrated
implementation provide a practical path toward more efficient
inter-organizational coordination. By enabling platform-independent rapid
consensus discovery and efficient treaty creation solutions, the framework
offers concrete mechanisms to transcend traditional diplomacy constraints,
aiming to enable humanity to act more like a unit and successfully navigate the
global challenges of the 21st century.

\vspace{12pt}
\bibliographystyle{plain}

\end{document}